# Latent Collaborative Retrieval

Jason Weston                                    JWESTON@GOOGLE.COM
Chong Wang                              CHONGW@CS.PRINCETON.EDU
Ron Weiss                                         RONW@GOOGLE.COM
Adam Berenzeig                              MADADAM@GOOGLE.COM
Google, 76 9th Avenue, New York, NY, 10011 USA

## Abstract

Retrieval tasks typically require a ranking of items given a query. Collaborative filtering tasks, on the other hand, learn to model user's preferences over items. In this paper we study the joint problem of recommending items to a *user* with respect to a given *query*, which is a surprisingly common task. This setup differs from the standard collaborative filtering one in that we are given a query × user × item tensor for training instead of the more traditional user × item matrix. Compared to document retrieval we do have a query, but we may or may not have content features (we will consider both cases) and we can also take account of the user's profile. We introduce a factorized model for this new task that optimizes the top-ranked items returned for the given query and user. We report empirical results where it outperforms several baselines.

## 1. Introduction

There exist today a growing number of applications that seamlessly blend the traditional tasks of retrieval and recommendation. For example, when users shop for a product online they are often recommended items that are similar to the item they are currently browsing. This is a retrieval problem using the currently browsed item as the query, however the user's profile (including other items they may have browsed, bought or reviewed) should be taken into account making it a personal recommendation problem as well. Another related task is that of automatic playlist creation in music players. The user can request the creation of a playlist of songs given a query (based for instance on a seed track, artist or genre) but the songs retrieved for the query should also be songs that the user likes given their known profile.

We call this class of problems *collaborative retrieval* tasks. To our knowledge these tasks have not been studied in depth, although there are several related areas which we will discuss later in the paper. Methods designed for this task need to combine both the retrieval and recommendation aspects of the problem into a single predictor. In a standard collaborative filtering (recommendation) setup, one is given a user × item matrix indicating the known relevance of the given item to a given user, but many elements of the matrix are unknown. On the other hand, In a typical retrieval task one is given, for each query, a list of relevant items that should be retrieved. Our task is the blend of the two, which is achieved by first building a tensor comprising of the query × user × item training data. Typically in a retrieval task, and sometimes in a recommendation task as well, one also has access to *content-based features* for the items, e.g. in document retrieval one has access to the words in the documents. Hence any algorithms designed for the collaborative retrieval task should potentially be able to take advantage of those features, too.

In this paper, we develop a novel learning algorithm for the collaborative retrieval task. We introduce a factorized model that optimizes the top-ranked items returned for the given query and user. We also generalize it to work on either the collaborative retrieval tensor only, or using content-based features as well. The rest of the paper is as follows. Section 2 describes the collaborative retrieval task and our method for solving it. Section 3 discusses prior work and connections to other areas. Finally, Section 4 reports empirical results where we show our method outperforms several reasonable baselines, and Section 5 concludes.

## 2. Method

**Latent Collaborative Retrieval** We define a scoring function for a given query, user and item:

$$f_{FULL}(q, u, d) = \mathbf{R}_{qud}$$

where $\mathbf{R}$ is a $|\mathcal{Q}| \times |\mathcal{U}| \times |\mathcal{D}|$ tensor, where $\mathcal{Q}$ is the (finite) set of possible queries, $\mathcal{U}$ is the set of users and $\mathcal{D}$ is the set of items. Any given element of the tensor is the "relevance score" of a given item with respect to a given query *and* a given user, where a high score corresponds to high relevance.

We are typically given $m$ training examples $\{(\mathbf{q}_i, \mathbf{u}_i, \mathbf{d}_i)\}_{i=1,\ldots,m} \in \{1, \ldots, |\mathcal{Q}|\} \times \{1, \ldots, |\mathcal{U}|\} \times \{1, \ldots, |\mathcal{D}|\}$ and outputs $\mathbf{y}_i \in \mathbb{R}, i = 1, \ldots, m$. Here, $(\mathbf{q}_i, \mathbf{u}_i, \mathbf{d}_i)$ can be used to index a particular element of $\mathbf{R}$ (i.e, a particular query, user and item) and $\mathbf{y}_i$ is the relevance score, for example based on implicit user clicks, activity or explicit user annotations. One could simply collate the training data to build a suitable tensor $\mathbf{R}$ and use that, but the problem is that the tensor would be *sparse* and hence for many queries, users and items no prediction would be made. For that reason, collaborative filtering has connections with *matrix completion*, and almost all approaches can be seen as estimating the unknown matrix from data. For instance, many approaches such as SVD or NMF (Lee & Seung, 2001), solve such tasks by optimizing the deviation (e.g. squared error) from the known elements of the matrix. However, for retrieval tasks, and even for many recommendation tasks, humans evaluate the performance of a method from the top $k$ results returned. Hence, precision or recall @ k measures are often appropriate. The method we propose in this paper thus has the following properties:

(i) We learn a ranking of items given a user and a query, thus blending retrieval and recommendation tasks into one model.

(ii) We learn model parameters for this task that attempt to optimize the performance at the top of the ranked list.

To fulfill property (i) we must model the combination of the users, queries and items during inference. We thus propose a model of the following form:

$$f(q, u, d) = \Phi_Q(q)^\top S^\top U_u T \Phi_D(d) + \Phi_U(u)^\top V^\top T \Phi_D(d). \quad (1)$$

Here, $S$ is a $n \times |\mathcal{Q}|$ matrix, $T$ is a $n \times |\mathcal{D}|$ matrix, $V$ is a $n \times |\mathcal{U}|$ matrix and $n$ is the low dimensional embedding where queries, users and items will be represented (this

is a hyperparameter of the system, typically $n \ll |\mathcal{D}|$ and $n \ll |\mathcal{U}|$). $U_i$ is a $n \times n$ matrix per user ($i = 1, \ldots, |\mathcal{U}|$). $\Phi_D(d)$ is the feature map of the item, the simplest choice of which is to map to a binary vector of all zeros and a one in the $d^{th}$ position. $\Phi_Q(q)$ and $\Phi_U(u)$ act similarly for queries and users. In that case, the entire model can hence be more succinctly written as:

$$f(q, u, d) = (S_q^\top U_u + V_u^\top) T_d. \quad (2)$$

However the $\Phi(\cdot)$ notation will be useful for subsequent modifications of the algorithm (later, we will consider general feature transformations rather than just switching on a single dimension).

An intuitive explanation of our model is as follows. The first term maps both the query (via $\Phi_Q(q)^\top S^\top$) and the item (via $T \Phi_D(d)$) into a low dimensional space and then measures their similarity in that space, after linearly transforming the space dependent on the user (via $U_u$). Hence, the first term alone can model the relevance score (match) between a query $q$ and item $i$ with respect to a user $u$. The second term can be seen as a kind of "bias" that models the relevance score (match) between user $u$ and item $i$ but is constant w.r.t the query.

It is also possible to consider some interesting special cases of the above model by further constraining the user-transformation matrices $U_i$:

- $U_i = I$: by forcing all user-transformations to be the identity matrix we are left with the model:

$$f(q, u, d) = S_q^\top T_d + V_u^\top T_d. \quad (3)$$

  In that case, the query $\times$ item and user $\times$ item parts of the model are two separate terms, and three-way interactions are not directly considered.

- $U_i = D_i$: by constraining each user $k$ to have a diagonal matrix $D_i$ only rescaling of the dimensions of the query $\times$ item similarity space is possible (general linear transformations are not considered). As we will see in Section 3.2 this relates to tensor factorization methods that have been used for document retrieval.

- $U_i = (U_i^{LR})^\top U_i^{LR} + D_i$: instead of considering a full matrix $U$ or a diagonal $D_i$ we could consider something in between, that of employing a low rank matrix $U^{LR}$.

**Content-Based Method** In the typical collaborative filtering setting one has access to a user $\times$ item matrix only, and methods are agnostic to the *content* of the items, be they text documents, audio or images.

For some tasks one has access to the actual content of the items as well, for example for each item $i$ one is given a feature representation $\hat{\Phi}_D(i) \in \mathbb{R}^{n_D}$. In document retrieval this is the more common setting, e.g. $\hat{\Phi}_D(i)$ represents the words in the document $i$. For recommendation tasks this is called content-based recommendation and is particularly useful for the *cold-start* problem where an item has very few or no users associated to it (the relevant collaborative filtering column of $R$ is very sparse). In that case, collaborative filtering methods have almost no data to generalize from, but content-based methods can perform well. In our setting, latent collaborative retrieval, we can also take advantage of such content features by slightly modifying our method from above. Our proposed content-based model consists of the following form:

$$f(q, u, d) = S_q^\top U_u W_D \hat{\Phi}_D(d) + V_u^\top W_D \hat{\Phi}_D(d). \quad (4)$$

Here, the model is similar to before except an additional set of parameters $W_D$ (a $n \times n_D$ matrix) maps from item *features* to the $n$-dimensional latent embedding space. Other aspects of the model remain the same.

Further, if we are given a feature representation for queries as well, where for each query $i$ we have $\hat{\Phi}_Q(i) \in \mathbb{R}^{n_Q}$, we can also incorporate this into our model:

$$f(q, u, d) = \hat{\Phi}_Q(q)^\top W_Q^\top U_u W_D \hat{\Phi}_D(d) + V_u^\top W_D \hat{\Phi}_D(d), \quad (5)$$

where $W_Q$ is a $n \times n_Q$ matrix. This allows us to consider any possible query rather than being restricted to a finite set $\mathcal{Q}$ as in our original definition.

**Collaborative and Content-Based Retrieval** Finally, we can consider a joint model that takes into account both collaborative filtering (CF) data and content-based (CB) training data. In this case, our model consists of the following form:

$$\begin{aligned} f(q, u, d) = {} & S_q^\top U_u W_D \hat{\Phi}_D(d) + S_q^\top U_u T_d + \\ & \hat{\Phi}_Q(q)^\top W_Q^\top U_u W_D \hat{\Phi}_D(d) + \hat{\Phi}_Q(q)^\top W_Q^\top U_u T_d + \\ & V_u^\top W_D \hat{\Phi}_D(d) + V_u^\top T_d. \end{aligned} \quad (6)$$

The first two terms match the query and user with the CF and CB versions of the item respectively. Terms three and four are similar except they use the content features of the query instead. The final two terms are the "bias" terms comparing the user to the CF and content versions of the item. Note this model can be considered a special case of eq. (1).

**Training To Optimize Retrieval For The Top $k$** We are interested in learning a *ranking* function where the top $k$ retrieved items are of particular interest as they will be presented to the user. We wish to optimize all the parameters of our model jointly for that goal.

A standard loss function is often used for retrieval is the margin ranking criterion (Herbrich et al., 2000; Joachims, 2002), in particular it was used for learning factorized document retrieval models in Bai et al. (2009). Let us first write the predictions of our model for *all* items in the database as a vector $\bar{f}(q, u)$ where the $i^{th}$ index is $\bar{f}_i(q, u) = f(q, u, i)$. In our collaborative retrieval setting the loss can then be written as:

$$err_{AUC} = \sum_{i=1}^m \sum_{j \neq \mathbf{d}_i} \max(0, 1 - \bar{f}_{\mathbf{d}_i}(q_i, u_i) + \bar{f}_j(q_i, u_i)). \quad (7)$$

For each training example $i = 1, \dots, m$, the positive item $\mathbf{d}_i$ in the example triplet is compared to all possible negative items $j \neq \mathbf{d}_i$, and one assigns to each pair a cost if the negative item is larger or within a "margin" of 1 from the positive item. These costs are called *pairwise violations*. Note that all pairwise violations are considered equally if they have the same margin violation, independent of their position in the list. For this reason the margin ranking loss might not optimize the top $k$ very accurately as it cares about the average rank.

To instead focus on the top of the ranked list of returned items we employ a recently introduced loss function that has been developed for document retrieval (Usunier et al., 2009; Weston et al., 2010; 2012). To the best of our knowledge this method has not been applied to collaborative filtering type tasks before. The main idea is to *weigh* the pairwise violations depending on their position in the ranked list. One considers a class of ranking error functions:

$$err_{WARP} = \sum_{i=1}^m L(rank_{\mathbf{d}_i}(\bar{f}(\mathbf{q}_i, \mathbf{u}_i))) \quad (8)$$

where $rank_{\mathbf{d}_i}(\bar{f}(q_i, u_i))$ is the margin-based rank of the labeled item given in the $i^{th}$ training example:

$$rank_i(\bar{f}(q, u)) = \sum_{j \neq i} \theta(1 + \bar{f}_j(q, u) \geq \bar{f}_i(q, u))$$

where $\theta$ is the indicator function, and $L(\cdot)$ transforms this rank into a loss:

$$L(r) = \sum_{i=1}^r \alpha_i, \text{ with } \alpha_1 \geq \alpha_2 \geq \cdots \geq 0. \quad (9)$$

Different choices of $\alpha$ define different weights (importance) of the relative position of the positive examples in the ranked list. In particular it was shown that by choosing $\alpha_i = 1/i$ a smooth weighting over positions is given, where most weight is given to the top position, with rapidly decaying weight for lower positions. This is useful when one wants to optimize precision at $k$ for a variety of different values of $k$ at once (Usunier et al., 2009). (Note that choosing $\alpha_i = 1$ for all $i$ we have the same AUC optimization as equation (7)).

We optimize this function by stochastic gradient descent (SGD) following the authors of (Weston et al., 2010), that is samples are drawn at random, and a gradient step is made for each draw. Due to the cost of computing the exact rank in (8) it is approximated by sampling. That is, for a given positive label, one draws negative labels until a violating pair is found, and then approximates the rank with

$$rank_d(\bar{f}(q, u)) \approx \left\lfloor \frac{|\mathcal{D}| - 1}{N} \right\rfloor$$

where $\lfloor . \rfloor$ is the floor function, $|\mathcal{D}|$ is the number of items in the database and $N$ is the number of trials in the sampling step. Intuitively, if we need to sample more negative items before we find a violator then the rank of the true item is likely to be small (i.e., at the top of the list, as few negatives are above it).

Finally, our models have many parameters to be learnt. One can regularize them by preferring smaller weights. We constrain the parameters using $||S_i|| \leq C$, $i = 1, \ldots, |\mathcal{Q}|$, $||V_i|| \leq C$, $i = 1, \ldots, |\mathcal{U}|$, $||T_i|| \leq C$, $i = 1, \ldots, |\mathcal{D}|$ (leaving $U$ unconstrained). During SGD one projects the parameters back on to the constraints at each step, following the same procedure used in several other works, e.g. (Weston et al., 2010; Bai et al., 2009).

## 3. Prior Work and Connections

### 3.1. Connections to matrix factorization for collaborative filtering

Many works for collaborative filtering tasks have proposed using factorized models. In particular, Singular Value Decomposition (SVD) and Non-negative Matrix Factorization (NMF) (Billsus & Pazzani, 1998; Lee & Seung, 2001) are two popular choices. The main two differences between our approach and these general matrix factorization techniques is that (i) each recommendation we make is seeded with a query (i.e. the collaborative retrieval task), and (ii) in collaborative retrieval tasks we are interested in the top $k$ returned items, so our method optimizes for that goal.

Most collaborative filtering work does not consider a ranking type loss that optimizes the top $k$, but one notable exception is (Weimer et al., 2007). They do not consider tensor factorizations.

There are several ways to factorize a tensor, some classical ways are Tucker decomposition (Tucker, 1966) and PARAFAC (Harshman, 1970). Several collaborative filtering techniques have considered tensor factorisations before, in particular for taking into account user context features like tags (Rendle & Schmidt-Thieme, 2010), web pages (Menon et al., 2011), age and gender (Karatzoglou et al., 2010), time (Xiong et al., 2010) or user location for mobile phone recommendation (Zheng et al., 2010) but not, to our knowledge, for the collaborative retrieval task.

Finally, we should also note that some works have combined collaborative filtering data with content-based features before, e.g. (Wang & Blei, 2011).

### 3.2. Connections to matrix factorization and information retrieval

In information retrieval one is required to rank items (documents) given a query using the content-features of the items, e.g. for document retrieval one uses the words in the document. In that case, Latent Semantic Indexing (Deerwester et al., 1990), and related methods such as LDA (Blei et al., 2003), are unsupervised methods that choose a low dimensional feature representation of the words. The parameterization of those models is a special case of our models. If we consider our model from equation (5) but remove the influence of the user model, i.e. set $U_u = I$ and $V_u = 0$ we are left with a standard document retrieval model:

$$f_{DR}(q, d) = \hat{\Phi}_Q(q)^\top W_Q^\top W_D \hat{\Phi}_D(d). \quad (10)$$

More recently, factorized models that are supervised to the task of document retrieval have been proposed, for example Polynomial Semantic Indexing (PSI) (Bai et al., 2009). PSI considers polynomial terms between document words and query words for higher order similarities. For degree 2 it has the form of (10) but for degree 3 it uses tensor factorizations based on:

$$f^3(q, d) = \sum_k (S\hat{\Phi}_Q(q))_k (U\hat{\Phi}_D(d))_k (V\hat{\Phi}_D(d))_k.$$

This is closely related to our model (5) when constraining $U_i = D_i$ and replacing the user input by the document input, i.e. we compute $f(q, d, d)$ in order to obtain interactions between document words rather than between document and user.

Methods like LSI or LDA optimize the reconstruction error (mean squared error or likelihood). PSI optimizes the AUC ranking loss, which is more related to our ranking approach but does not optimize the top $k$ results like ours. Methods for annotating images (Weston et al., 2010) and labeling songs with tags (Weston et al., 2012) have been proposed that do use the WARP loss we employ in this paper. Many methods for document retrieval also optimize the top $k$ but typically not using factorized models like ours, see e.g. (Yue et al., 2007).

Finally, our models are applicable to the task of "personalized search" where some topic model approaches have recently been studied (Harvey et al., 2011; Lin et al., 2005; Sun et al., 2005; Saha et al., 2009). We will compare to generalized SVD and NMF models in our experiments which are related to these works.

## 4. Experiments

Traditional collaborative filtering datasets like the Netflix challenge dataset, and information retrieval datasets, like LETOR for instance, cannot be used in the collaborative retrieval framework, as they either lack the query or the user information necessary. We therefore use the three datasets described below.

### 4.1. Lastfm Dataset

We used the "Last.fm Dataset - 1K users" dataset available from http://www.dtic.upf.edu/~ocelma/MusicRecommendationDataset/lastfm-1K.html. This dataset contains (user, timestamp, artist, song) tuples collected from the Last.fm (www.lastfm.com) API. This dataset represents the listening history (until May 5th, 2009) for 992 users and 176,948 artists. Two consecutively played artists by the same user are considered as a query × user × item triple. This mirrors the task of playlisting, where a user selects a seed track, and the machine has to automatically build a list of tracks. We consider two artists as "consecutive" if they are played within an hour of each other (via the timestamp), otherwise we ignore the pair. One in every five days (so that the data is disjoint) were left aside for testing, and the remaining data was used for training and validation. Overall this gave 5,408,975 training triples, 500,000 validation triples and 1,434,568 test triples.

### 4.2. Playlist head and tail datasets

We had access to a larger scale proprietary and anonymized dataset of user playlists where we could both construct a query × user × item matrix from consecutive tracks, and had access to content-based features as well so we can test our content-based feature methods. The first extracted dataset ("head" dataset) consists of 46,000 users and 943,284 tracks from 146,369 artists (each artist appears at least 10 times). The data is split into 17M training triples for training, 172,000 for validation and 1.7M for test.

The above "head" dataset can be built for artists where we have enough training data. However, a user may want to do retrieval with a query or an item for which we have no tensor training data at all (i.e., the cold-start problem). In that case, content-based feature approaches are the only option. To evaluate this setup we hence built a "tail" testing dataset consisting of 10,000 triples from 5442 artists where we only have a single test example. The idea in that case is to train on the head dataset, and test on the tail (as it is not possible to train on the tail).

For each track (including head tracks) we have access to the audio features of the track, which we processed using the well-known Mel Frequency Cepstral Coefficient (MFCC) representation. MFCCs take advantage of source/filter deconvolution from the cepstral transform and perceptually-realistic compression of spectra from the Mel pitch scale and have been used widely in music and speech (Foote, 1997; Rabiner & Juang, 1993). We extracted 13 MFCCs every 10ms over a Hamming window of 25ms, and first and second derivatives were concatenated, for a total of 39 features. We then computed a dictionary of 2000 typical MFCC vectors over the training set (using K-means) and represented each song as a vector of counts, over the set of frames in the given song, of the number of times each dictionary vector was nearest to the frame in the MFCC space. The resulting feature vectors thus have dimension $n_D = 2000$.

### 4.3. Baselines

We compare to Singular Value Decomposition (SVD) and Non-negative Matrix Factorization (NMF) which are both popular methods for collaborative filtering tasks. For SVD we use the Matlab implementation and for NMF we use the implementation at http://www.csie.ntu.edu.tw/~cjlin/nmf/. Standard SVD and NMF operate on matrices, not tensors, so we compare our method on those tasks (where we consider only user × item matrices or only query × item matrices) as well. For the query × user × item tensor we considered the following generalization of SVD or NMF:

$$f(q, u, i) = \Phi(q)^\top U_{QI}^\top V_{QI} \Phi(d) + \gamma \Phi(u)^\top U_{UI}^\top V_{UI} \Phi(d).$$

That is, we perform two SVDs (or NMFs), one for the user × item matrix and one for the query × item matrix, and then combine them with a mixing parameter $\gamma$ which is chosen on the validation set.

For LCR, we compare both the versions from equation

*Table 1.* Recommendation Results on the LASTFM dataset.

| METHOD | R@5 | R@10 | R@30 | R@50 |
|---|---|---|---|---|
| NMF QUERY X ITEM | 3.76% | 6.38% | 13.3% | 17.8% |
| SVD QUERY X ITEM | 4.01% | 6.93% | 13.9% | 18.5% |
| LCR QUERY X ITEM | 5.60% | 9.49% | 18.9% | 24.8% |
| NMF USER X ITEM | 6.05% | 9.86% | 20.3% | 26.5% |
| SVD USER X ITEM | 6.60% | 10.7% | 21.4% | 27.7% |
| LCR USER X ITEM | 8.37% | 14.0% | 27.8% | 36.5% |
| NMF QUERY X ITEM + USER X ITEM | 5.96% | 9.93% | 20.5% | 26.6% |
| SVD QUERY X ITEM + USER X ITEM | 6.82% | 12.1% | 25.9% | 34.9% |
| LCR QUERY X ITEM + USER X ITEM | 9.22% | 15.1% | 30.2% | 39.0% |
| LCR QUERY X USER X ITEM | 10.6% | 16.6% | 32.2% | 41.2% |

*Table 2.* Optimizing r@*k* (WARP) versus optimizing AUC.

| METHOD | AUC R@10 | AUC R@30 | WARP R@10 | WARP R@30 |
|---|---|---|---|---|
| LCR QUERY X ITEM | 6.32% | 14.8% | 9.49% | 18.9% |
| LCR USER X ITEM | 11.0% | 23.7% | 14.0% | 27.8% |
| LCR QUERY X ITEM + USER X ITEM | 12.1% | 25.9% | 15.1% | 30.2% |

(3) and equation (2) on the query × user × item task. The former is directly comparable to the SVD and NMF tensor generalizations we use (they are the same paramaterization) while the latter takes into account three-way interactions between query, user and item in a single joint formulation. For the user × item and query × item tasks we employ either only the first or the second term respectively of equation (3). The validation set is used to choose the hyperparameters, e.g. the best choice of learning rate, regularization parameter and as a stopping criterion for the gradient descent.

For content-based features we also compare to LSI (Deerwester et al., 1990) and using cosine similarity. Both of these methods perform retrieval given the query, and ignore the user term.

### 4.4. Evaluation

For any given query $q$, user $u$, item $i$ triple we compute $f(q, u, \hat{i})$ using the given algorithm for each possible item $\hat{i}$ and sort them, largest first. For user × item or query × item tasks the setup is the same except either $q$ or $u$ is not used in all the competing models. The evaluation score for a given triple is then computed according to where item $i$ appears in the ranked list. We measure recall@k, which is 1 if item $i$ appears in the top $k$, and 0 otherwise. We report mean recall@k over the entire test set. Note that as we only consider one positive example per query (the element $i$ of the triple) precision@k = recall@k / $k$.

### 4.5. LastFM dataset Results

We first report results on the LASTFM dataset. Detailed results where we fixed the embedding dimension of all methods to $n = 50$ are given in Table 1. Results for other choices of $n$ are given in Table 4. On all three tasks (query × item, user × item and query × item) LCR is superior to SVD and NMF for each top-ranked set $k$ considered. Furthermore, our full LCR query × user × item model (c.f. equation 2, $U_i$ unconstrained) gives improved results compared to both (i) any competing methods, including LCR itself, that do not take into account both query *and user*; and (ii) LCR itself (and other methods) that do not model the query and user in a joint similarity function (i.e. LCR query x user x item (cf. eq. (2)) outperforms LCR query x item + user x item (cf. eq. (3))).

**Loss function evaluation** Some of the improvement of LCR over SVD and NMF can be explained by the fact that neither SVD nor NMF optimize a ranking function that optimizes the top-ranked items. To show the importance of the loss function, we report the results of LCR using an alternative loss function optimizing average rank (AUC) as in equation (7) instead of the WARP loss from equation (8). The comparison, given in Table 2 shows a clear gain on all tasks by optimizing for the top $k$ (using WARP). Optimizing AUC instead yields results in fact similar to SVD. SVD optimizes mean squared error, not AUC, but the similarity is that neither loss function pays special attention to the top $k$ results.

*Table 3.* Recommendation Results on the PLAYLIST head dataset.

| METHOD | R@5 | R@10 | R@30 | R@50 |
|---|---|---|---|---|
| NMF QUERY X ITEM | 5.28% | 8.87% | 18.4% | 24.0% |
| SVD QUERY X ITEM | 7.21% | 11.0% | 20.8% | 26.9% |
| LCR QUERY X ITEM | 10.7% | 16.3% | 29.1% | 35.6% |
| | | | | |
| NMF USER X ITEM | 6.23% | 10.2% | 19.2% | 25.0% |
| SVD USER X ITEM | 6.84% | 11.2% | 20.9% | 26.9% |
| LCR USER X ITEM | 6.26% | 10.5% | 21.8% | 29.3% |
| | | | | |
| NMF QUERY X ITEM + USER X ITEM | 6.26% | 10.2% | 19.2% | 25.0% |
| SVD QUERY X ITEM + USER X ITEM | 7.87% | 12.0% | 22.2% | 28.4% |
| LCR QUERY X ITEM + USER X ITEM | 12.8% | 19.4% | 34.5% | 42.0% |
| LCR QUERY X USER X ITEM | 13.0% | 19.6% | 34.6% | 42.2% |

*Table 4.* Changing the embedding size on the LASTFM dataset. We report R@30 for various dimensions $n$.

| METHOD | $n =$ | 10 | 25 | 50 | 100 |
|---|---|---|---|---|---|
| NMF QUERY X ITEM | | 8.53% | 9.94% | 13.3% | 12.3% |
| SVD QUERY X ITEM | | 11.5% | 12.8% | 13.9% | 14.7% |
| LCR QUERY X ITEM | | 15.0% | 18.0% | 18.9% | 19.8% |
| | | | | | |
| NMF USER X ITEM | | 12.1% | 16.5% | 20.3% | 23.5% |
| SVD USER X ITEM | | 12.8% | 17.7% | 21.4% | 25.1% |
| LCR USER X ITEM | | 21.1% | 26.1% | 27.8% | 28.7% |
| | | | | | |
| NMF Q X I + U X I | | 12.7% | 16.9% | 20.5% | 23.6% |
| SVD Q X I + U X I | | 13.3% | 17.9% | 25.9% | 25.7% |
| LCR Q X I + U X I | | 22.3% | 27.9% | 30.2% | 31.3% |
| LCR QUERY X USER X ITEM | | 23.3% | 28.9% | 32.2% | 33.3% |

**Changing the embedding dimension** We report results varying the embedding dimension $n$ in Table 4. It should be noted that $n$ affects both test performance, evaluation time and storage requirements, so low dimensional embeddings are preferable if they perform well enough. LCR outperforms the baselines for all values of $n$ that we tried, however all methods degrade significantly when $n = 10$. SVD on the query × user × item shows the same performance for $n = 50$ and $n = 100$ while LCR improves slightly.

### 4.6. Playlist dataset results

**Collaborative filtering type data** The PLAYLIST dataset is larger scale and has both collaborative-filtering type data and content-based features. We first tested using collaborative filtering type data only on the same three tasks as before (query × item, user × item and query × user × item). The results are given in Table 3. They again show a performance improvement for LCR over the SVD and NMF baselines on the query × item task, although on the user × item task it performs similarly to the baselines. However, on the most interesting task, query × user × item, we again see a large performance gain.

**Using content-based features** We compared different algorithms using content-based features on the tail dataset where collaborative filtering cannot be used. (We also attempted to combine both collaborative filtering and content-based information on the head dataset, but we observed no gain in performance over collaborative filtering alone, probably because the content-based features are not strong enough, which is not really a surprising result (Slaney, 2011)). The results on the tail dataset are given in Table 5. LCR q×i (which does not use user information, as in eq. (10)) already outperforms cosine similarity and LSI. Adding user information further improves performance: LCR q×u+q×i uses the model form of eq. (4) with $U_i = I$ and LCR query × user × item uses eq. (4) with $U_i = D_i$.

## 5. Conclusion

In this paper we introduced a new learning framework called *collaborative retrieval* which links the standard document retrieval and collaborative filtering tasks. Like collaborative filtering, the task is to rank items given a user, but crucially we can also take into ac-count a query term. Like document retrieval we are given a query and the task is to rank items, but crucially we also take into account the user in the form of a user × query × item tensor of training data.

*Table 5.* Content-based Results on the PLAYLIST tail set.

| METHOD | R@30 | R@50 | R@100 | R@200 |
|---|---|---|---|---|
| COSINE | 1.1% | 1.7% | 2.9% | 5.2% |
| LSI | 1.0% | 1.5% | 2.6% | 4.8% |
| LCR QUERY×ITEM | 1.9% | 2.9% | 4.8% | 8.4% |
| LCR Q×I+U×I | 2.3% | 3.4% | 5.9% | 10.3% |
| LCR QUERY×USER×ITEM | 2.4% | 3.5% | 6.0% | 10.7% |

We proposed a novel learning algorithm for this task that learns a factorized model to rank the items given the query and user, and showed it empirically outperforms some standard methods. Collaborative retrieval is rapidly becoming an important task and we expect this to become a well studied research area.